\documentclass{ws-mplb}
\usepackage{amsfonts,amssymb,amsmath}
\usepackage{graphicx}
\usepackage[super,sort,compress]{cite} 

\begin{document}


%
%

\title{PREPARATION OF A SUPERPOSITION OF SQUEEZED COHERENT STATES OF A CAVITY FIELD VIA COUPLING TO A SUPERCONDUCTING CHARGE QUBIT}

\author{DAGOBERTO S. FREITAS}

\address{Departamento de F\'\i sica, Universidade
Estadual de Feira de Santana, \\
Feira de Santana, Bahia, 44036-900, Brazil\\
dfreitas@uefs.br}



\maketitle


\begin{abstract}
The generation of nonclassical states of a radiation field has
become increasingly important in the past years given its various
applications in quantum communication. The feasibility of generating
such nonclassical states has been established in several branches of
physics such as cavity electrodynamics, trapped ions, quantum dots,
atoms inside cavities and so on. In this sense, we will discuss the
issue of the generation of nonclassical states in the context of a
superconducting qubit in a microcavity. It has been recently
proposed a way to engineer quantum states using a SQUID charge qubit
inside a cavity with a controllable interaction between the cavity
field and the charge qubit. The key ingredients to engineer these
quantum states are a tunable gate voltage and a classical magnetic
field applied to SQUID. In Ref. [Yu-xi Liu, L. F. Wei, and F. Nori,
Europhys. Lett., 67, 941 (2004)] it was proposed a model including
these ingredients and using some appropriate approximations which
allow for the linearization of the interaction and nonclassical
states of the field were generated. Since decoherence is known to
affect quantum effects uninterruptedly and decoherence process are
works even when the quantum state is being formed, therefore, it is
interesting to envisage processes through which quantum
superpositions are generated as fast as possible. The decoherence
effect has been studied and quantified in the context of cavity QED
where it is shown that the more quantum is the superposition, more
rapidly the environmental effects occur during the process of
creating the quantum state. In a previous work we have showed that
the contribution of nonlinear interaction is essential to shorten
the time necessary to build the quantum state and we have presented
an approach for preparing a Schr\"{o}dinger cat (SC) states of mode
of cavity field interacting with a superconducting charge qubit [D.
S. Freitas, and M. C. Nemes, Modern Physics Letters B, 28, 10
(2014)]. In the latter reference, we have succeeded in linearizing
the Hamiltonian through the application of an appropriate unitary
transformation and for certain values of the parameters involved, we
have showed that it is possible to obtain specific Hamiltonians. In
this work we will use such approach for preparing superposition of
two squeezed coherent states.
\end{abstract}


\section{Introduction}
In the past years, given its applications in quantum communication,
the generation of nonclassical states of a radiation field has
become more and more important. The possibility of generating
nonclassical states has been possible in various branches of
physics, such as cavity electrodynamics, trapped ions, quantum dots,
atoms within cavities, and so on \cite{brune96,monroe96}. In this
sense, we will discuss the generation of non-classical states in the
context of a superconducting qubit in a microcavity. Recently, it
was proposed a way to project quantum states using a qubit SQUID
charge qubit inside a cavity with a controllable interaction between
the cavity field and the charge qubit. The main ingredients for
projecting quantum states are a tunable gate voltage and a classic
magnetic field applied to SQUID. The recent interest in the study of
the cavity quantum electrodynamics type systems such as a
superconducting qubit can open new ways for studying the interaction
between light and solid-state quantum devices
\cite{buluta11,xiang13}. Various theoretical and experimental works
have discussed the interaction between superconducting qubits with
either quantized
\cite{yupra05,yu04,you03,jie07,wen08,song09,yong11,liao11,zago04,mahmoud09,ashhab10,ke10,valverde11,valverde12,tang14}
or classical fields \cite{zhou02,paspa04,yuprl05}. Recently, it has
been proposed a way to project quantum states using a
superconducting quantum device (SQUID) charge qubit inside a cavity
\cite{yupra05,yu04} with a controllable interaction between the
cavity field and the charge qubit. In references \cite{yupra05,yu04}
the model proposed including these ingredients and using some
adequate approximations which allow for the linearization of the
interaction and nonclassical states of the field are generated. The
single-cavity scheme of reference \cite{yupra05,yu04} may be
extended to generate entangled coherent states of two microwave
cavity fields coupled to a SQUID - type superconducting box, as
proposed in reference \cite{jie07}. In the literature it can also be
found proposals for the generation of entangled states and squeezed
states using linear and nonlinear interactions between microwave
cavity field and SQUID - type superconducting box \cite{wen08};
schemes for generation of multi-qubit entangled cluster states
\cite{song09}, for deterministic generation of entangled photon
pairs in a superconducting resonator array \cite{yong11}, and for
controlling the entanglement between two Josephson charge qubits
\cite{liao11}. We show that the essential contribution of the
nonlinear interaction is to shorten the time necessary to build the
quantum state. Since decoherence is known to affect quantum effects
uninterruptedly, they are at work even while the quantum state is
being formed. This has been studied and quantified in the context of
cavity QED where it is shown that the more "quantum" is the
superposition the more rapid are the environmental effects during
the process of creating the quantum state \cite{ze02}. It is
therefore interesting to envisage processes through which quantum
superpositions are generated as fast as possible.

\section{The model}
\begin{figure}[th]
\centerline{\includegraphics[width=7cm]{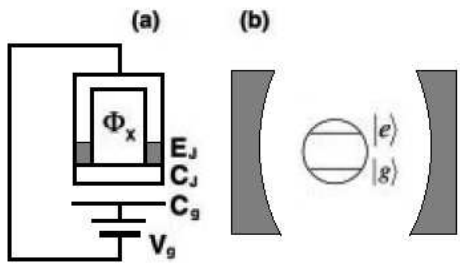}} \vspace*{8pt}
\caption{(a) A charge qubit formed by a SQUID device, equivalent to
a controllable macroscopic two-level system, is placed into a
superconducting microwave cavity in (b). The coupling between the
quantized cavity field and qubit system is realized via the magnetic
flux $\Phi_{X}$ through the SQUID. Figure adapted from Yu-xi Liu, L.
F. Wei, and F. Nori, {\it Europhys. Lett.} {\bf 67} (2004)
941.\label{figure}}
\end{figure}

We consider a system constituted by a SQUID type superconducting box
with $n_{c}$ excess Cooper-pair charges connected to a
superconducting loop via two identical Josephson junctions having
capacitors $C_{J}$ and coupling energies $E_{J}$, see
Fig.(\ref{figure}a). An external control voltage $V_{g}$ couples to
the box via a capacitor $C_{g}$. We also assume that the system
operates in a regime consistent with most experiments involving
charge qubits, in which only Cooper pairs coherently tunnel in the
superconducting junctions. Therefore, the system Hamiltonian may be
written as \cite{yupra05,yu04,yu01}
\begin{equation}
H_{qb} = 4E_{ch}(n_{c}-n_{g})^{2}-2E_{J}\cos(\frac{\pi
\Phi_{X}}{\Phi_{0}})\cos(\Theta),\label{Hqb}
\end{equation}
where $E_{ch} = e^{2}/2(C_{g} +2C_{J})$ is the single-electron
charging energy, $n_{g} = C_{g}V_{g}/2e$ is the dimensionless gate
charge (controlled by $V_{g}$), $\Phi_{X}$ is the total flux through
the SQUID loop and $\Phi_{0}$ the quantum flux. The phase $\Theta$
is the quantum-mechanical conjugate of the number operator $n_{c}$
of the Cooper pairs in the box. The superconducting box is assumed
to be working in the charging regime and the superconducting energy
gap $\Delta$ is considered to be the largest energy involved.
Moreover, the temperature $T$ is low enough so that $\Delta \gg
E_{ch} \gg E_{J} \gg k_{B}T$, where $k_{B}$ is the Boltzmann
constant.The superconducting box then becomes a two-level system
with states $|g\rangle$ (for $n_{c} = 0$) and $|e\rangle$ (for
$n_{c} = 1$) given that the gate voltage is near a degeneracy point
($n_{g} = 1/2$) \cite{yu01} and the quasi-particle excitation is
completely suppressed \cite{averin}.

If the circuit is placed within a single-mode microwave
superconducting cavity, the qubit can be coupled to both a classical
magnetic field (generates a flux $\Phi_{c}$) and the quantized
cavity field (generates a flux $\Phi_{q} =\eta a +\eta^{*}
a^{\dag}$, with $a$ and $a^{\dag}$ the annihilation and creation
operators), being the total flux through the SQUID given by
$\Phi_{X} = \Phi_{c} + \Phi_{q}$ \cite{you03}, see
Fig.(\ref{figure}). The parameter $\eta$ is related to the mode
function of the cavity field. The Hamiltonian system will then read
\begin{equation}
H = \hbar\omega a^{\dag}a + E_{z}\sigma_{z} -
E_{J}\sigma_{x}\cos\big(\gamma I + \beta a +\beta^{*}
a^{\dag}\big),\label{H}
\end{equation}
where we have defined the parameters $\gamma = \pi\Phi_{c}/\Phi_{0}$
and $\beta = \pi\eta/\Phi_{0}$. The first term corresponds to the
free cavity field with frequency $\omega = 4E_{ch}/\hbar$ and the
second one to the qubit having energy $E_{z} = -2E_{ch}(1 - 2n_{g})$
with $\sigma_z$ and $\sigma_x$ the Pauli matrices. The third term is
the (nonlinear) photon-qubit interaction term which may be
controlled by the classical flux $\Phi_{c}$. In general the
Hamiltonian in equation (\ref{H}) is linearized under some kind of
assumption. In Ref. \cite{yupra05,yu04}, for instance, the authors
decomposed the cosine in Eq.(\ref{H}) and expanded the terms
$\sin[\pi(\eta\, a+H.c.)/\Phi_{0}]$ and $\cos[\pi(\eta\,
a+H.c.)/\Phi_{0}]$ as power series in $a \,(a^{\dagger})$. In this
way, if the limit $|\beta|\ll 1$ is taken, only single-photon
transition terms in the expansion are kept, and  a Jaynes-Cummings
type Hamiltonian (JCM) is then obtained. In contrast to that, in the
reference \cite {moya,feng01,dago14,dago12,moya16} it is presented a
technique that obtain a JCM Hamiltonian valid for any value of $
|\beta|$. This technique consists in applying a unitary
transformation that linearizes the Hamiltonian of the system. After
transforming the original Hamiltonian, it is possible to obtain a
simpler Hamiltonian under certain resonance regimes. In other words,
it is possible linearize the superconductor/quantized field
Hamiltonian without doing the usual power series expansions of the
Hamiltonian.
\section{Dynamics of the system}
The central idea of the approach proposed in reference \cite{moya}
which allows for the inclusion of nonlinear effects is the
following: a unitary transformation is constructed in a way that
diagonalizes the Hamiltonian leading it to a much simpler form. The
nonlinear effects are therefore guarded in the transformation
affecting directly the time evolution of the system in a tractable
manner. The comparison of our proposal with other method is not
simple. Normally the full hamiltonian is truncated after some kind
of approximation - for instance, by taking the limit $|\beta|<<1$, a
simple linearized \cite{yu04} or nonlinear Hamiltonians \cite{you03}
are obtained. In the method used here it is possible to obtain in a
direct way, a Hamiltonian which allows an exact solution for the
state vector in a specific resonance regime (as long as $|\beta|$).
However, as the nonlinear effects are somehow guarded in the
transformed Hamiltonian, they may give rise to a more complex
dynamics, for example: in reference \cite{dago14} for the
preparation a Schr\"{o}dinger cat (SC) of mode of cavity field
interacting with a superconducting charge qubit; in reference
\cite{dago12}, the resulting dynamics exhibits typical behavior of a
driven Jaynes-Cummings model \cite{hmcsmd94} (or a trapped ion
within a cavity \cite{moya}), but without the presence of a
classical driving field; in reference \cite{feng01} for the
preparation of {\bf SC} with cold ions. We believe that the approach
used here could be useful not only for establishing a direct
connection to other well-known models in quantum optics, but also
the exploration of different regimes in superconducting systems.
Next we apply a unitary transformation to the full Hamiltonian given
by (\ref{H}) and make approximations afterwards. By applying the
unitary transformation \cite{moya}
\begin{eqnarray}
T &=& \frac{1}{\sqrt{2}} \Big\{-\frac{1}{2} \Big[
 D^{\dag}\Big(\alpha,\gamma \Big) -  D\Big(\alpha,\gamma \Big)
\Big]I-\frac{1}{2} \Big[
 D^{\dag}\Big(\alpha,\gamma\Big) + D\Big(\alpha,\gamma
\Big) \Big] \sigma_z\nonumber\\
&+& D\Big(\alpha,\gamma \Big)\sigma_{+} + D^{\dag}\Big(\alpha,\gamma
\Big) \sigma_{-} \Big\} \label{T}
\end{eqnarray}
to the Hamiltonian in equation (\ref{H}), with
$D(\alpha,\gamma)=D(\alpha)e^{i\frac{\gamma}{2}}$ where $D(\alpha)
=\exp[(\alpha a^{\dag} -\alpha^{*}a)]$ is the Glauber's displacement
operator, with $\alpha = i\beta^{*}/2$, we obtain the following
transformed Hamiltonian
\begin{eqnarray}
H_{T}  & \equiv &  THT^{\dagger}\nonumber\\
& = & \hbar\omega
a^{\dag}a+\frac{E_{J}}{2}\sigma_{z}+i\frac{\hbar}{2}\big[\omega\big(\beta
a- \beta^{*}a^{\dag}\big) +
2i\frac{E_{z}}{\hbar}\big]\sigma_{x}\nonumber\\
& + & \frac{E_{J}}{2}\cos\big[2\big(\beta a+
\beta^{*}a^{\dag}\big)+2\gamma
\big]\sigma_{z}\nonumber\\
&-& i\frac{E_{J}}{2}\sin\big[2\big(\beta a+
\beta^{*}a^{\dag}\big)+2\gamma
\big]\big(\sigma_{+}-\sigma_{-}\big)+|\frac{\beta}{2}|^{2}.\label{HT0}
\end{eqnarray}
This result holds for any value of the parameter $\beta$. In the
regime in which $\hbar\omega|\beta| = 4|\beta|E_{ch} \gg E_{J}$,
that can be obtained for $|\beta| \geq 0.25$, the Hamiltonian in
Eq.(\ref{HT0}) becomes
\begin{eqnarray}
H_{T}  & \cong & \hbar\omega
a^{\dag}a+\frac{E_{J}}{2}\sigma_{z}+i\frac{\hbar\omega}{2}\big[\big(\beta
a- \beta^{*}a^{\dag}\big) +
2i\frac{E_{z}}{\hbar\omega}\big]\sigma_{x}.\label{HT1}
\end{eqnarray}
Our Hamiltonian in Eq.(\ref{HT1}) becomes a Jaynes-Cummings type
Hamiltonian. For $|\beta| = 0.25$ the charge regime, $E_{ch} \gg
E_{J}$, is satisfied. Note that in the approach of reference
\cite{yupra05}, the condition $|\beta| \ll 1$ is also necessary, but
for a different reason, i.e., to truncate the co-sine (sine) series.
We should remark that in our scheme the Jaynes-Cummings evolution
takes place in the transformed frame, differently from the model
developed in \cite{yupra05}. The term $|\frac{\beta}{2}|^{2}$ was
not taken into account because it just represents an overall phase.
The same setup and transformation given by (\ref{T}) may also be
employed (see reference \cite{dago14}) in a scheme for preparation
of superpositions of coherent states of a single-mode cavity field
(Schr\"{o}dinger cats - {\bf SC}) extending the approach of
Ref.\cite{yupra05}. The results is very similar to the {\bf SC}
obtained in Ref. \cite{yupra05}, but  in contrast to that we did not
use the condition $|\beta| \ll 1$. In our scheme, as $|\beta|$ is
large and the value of the amplitude of coherent states are
proportional to $t^2$, the time for preparing of an observable {\bf
SC} state is much shorter than that in other schemes.
\section{Squeezed Coherent State}
Now we show how to prepare a superposition of two squeezed coherent
states. To obtain this superposition we set $E_{z} = 0$ ( $n_{g} =
1/2$) in (\ref{HT1}). Here, we take advantage of the fact that the
Hamiltonian given in (\ref{HT1}) has not been approximated and,
therefore, there are no restriction on the values of their
parameters.  By transforming the Hamiltonian (\ref{HT1}) with the
unitary operators,
\begin{equation}
U_{1} = e^{\varepsilon_1(a^\dag\sigma_{+}+\sigma_{-}a)}
\end{equation}
\begin{equation}
U_{2} = e^{\varepsilon_2(a\sigma_{+}+\sigma_{-}a^\dag)}
\end{equation}
with $\varepsilon_1$, $\varepsilon_2 \ll 1$, and $\sigma_{-}$,
$\sigma_{+}$ are the Pauli matrices. Setting as
\begin{equation}
\varepsilon_1 = \frac{i\hbar\omega\beta}{2(E_{J}-\hbar\omega)}
\end{equation}
\begin{equation}
\varepsilon_2 = -\frac{i\hbar\omega\beta}{2(E_{J}+\hbar\omega)},
\end{equation}
where we consider $\beta$ as real. Remaining up to first order in
the expansion $e^{\varepsilon A}Be^{-\varepsilon A} = B+\varepsilon
[A,B]+\frac{\varepsilon^{2}}{2!}[A,[A,B]]+\cdots\thickapprox
B+\varepsilon [A,B]$, i.e., doing a small rotation, we obtain the
Hamiltonian
\begin{eqnarray}
H_{eff} & \equiv &
U_{2}U_{1}H_{T}U^{\dag}_{1}U^{\dag}_{2}\nonumber\\
& = & \hbar\omega
a^{\dag}a+\frac{E_{J}}{2}\sigma_{z}-\frac{(E_{J}-\hbar\omega)\beta^2\hbar^2\omega^2}{4(E_{J}^2-\hbar^2\omega^2)}(a^2
+ a^{\dag
2})\sigma_{z}\nonumber\\
&+&\frac{E_{J}\beta^2\hbar^2\omega^2}{E_{J}^2-\hbar^2\omega^2}(a^{\dag}a
+ \frac{1}{2})\sigma_{z}. \label{Heff1}
\end{eqnarray}
In the Eq.(\ref{Heff1}) the first interaction term describe a
squeezed state Hamiltonian and the second interaction one describe a
dispersive Hamiltonian.  For the regime in which $\hbar\omega-E_{J}
\gg 4E_{J}$ the Hamiltonian in Eq.(\ref{Heff1}) becomes a squeezed
Hamiltonian
\begin{eqnarray}
H_{eff}  & = & \hbar\omega
a^{\dag}a+\frac{E_{J}}{2}\sigma_{z}-\frac{(E_{J}-\hbar\omega)\beta^2\hbar^2\omega^2}{4(E_{J}^2-\hbar^2\omega^2)}(a^2
+ a^{\dag 2})\sigma_{z}. \label{Heff2}
\end{eqnarray}
Now to obtain a superposition of squeezed coherent states we will
make a rotation on Eq.(\ref{Heff2}) so that $\sigma_{z} \rightarrow
\sigma_{x}$. This rotation is equivalent to applying the operator
\begin{equation}
U_{R} = \exp(-i\frac{\pi}{4}\sigma_{y}) = \frac{\sqrt{2}}{2}\big[1 -
(\sigma_{+} - \sigma_{-})\big]
\end{equation}
on (\ref{Heff2}) obtain the Hamiltonian
\begin{eqnarray}
H_{SS}  & \equiv & U_{R}H_{eff}U^{\dag}_{R} =  \hbar\omega
a^{\dag}a+\frac{E_{J}}{2}\sigma_{x}-\frac{(E_{J}-\hbar\omega)\beta^2\hbar^2\omega^2}{4(E_{J}^2-\hbar^2\omega^2)}(a^2
+ a^{\dag 2})\sigma_{x}, \label{HSS}
\end{eqnarray}
with $U_{R}\sigma_{z}U^{\dag}_{R} = \sigma_{x}$. If the system is
initially in the coherent state $|\gamma\rangle$ and the charge
qubit is in the ground state $|g\rangle =
\frac{1}{\sqrt{2}}(|+\rangle - |-\rangle)$ where $|+\rangle$
($|-\rangle$) is eigenstate of the Pauli operator $\sigma_x$ with
the eigenvalue $1(-1)$, we can entangle qubit states with
superpositions of two different squeezed coherent states ({\bf SS})
evolving in time as
\begin{eqnarray}
|\Psi_{T}^{R}(t)\rangle  =
\frac{1}{\sqrt{2}}\Big[|\Phi(t)_{-}\rangle |e\rangle +
|\Phi(t)_{+}\rangle|g\rangle \Big],\label{PsR}
\end{eqnarray}
with the squeezed coherent states
\begin{eqnarray}
|\Phi(t)_{\pm}\rangle & = &
\frac{1}{\sqrt{2}}\Big[e^{-i\frac{E_{J}}{2}t}|\gamma,
-i\xi^2t/\hbar\rangle \pm e^{i\frac{E_{J}}{2}t}|\gamma,
i\xi^2t/\hbar\rangle\Big]
\end{eqnarray}
where $\xi^2 = \frac{\beta^2\hbar^2\omega^2}{4(E_{J}+\hbar\omega)}$,
$|\gamma, \mp i\xi^2t/\hbar\rangle = e^{-i\omega a^{\dag}at \mp
i\xi^{2}(a^2 + a^{\dag 2})t/\hbar}|\gamma\rangle$.
Here, $|\gamma, \mp i\xi^2t/\hbar\rangle$ denote squeezed coherent
states, and the degree of squeezing is determined by the
time-dependent parameter $\xi^2t/\hbar =
\frac{\beta^2\hbar\omega^2}{4(E_{J}+\hbar\omega)}t$.

The result is an entangled state involving qubit and a cavity field.
If one measures the charge state (either in $|g\rangle$ or
$|e\rangle$), the action will collapse the $|\Psi_{T}^{R}(t)\rangle$
into a {\bf SS} state $|\Phi_{\pm}\rangle$. The form of
Eq.(\ref{PsR}) is very similar to the {\bf SS} obtained in Ref.
\cite{yupra05}. But, in contrast to that, we did not do use the
condition $\beta \ll 1$. In our scheme, as $\beta$ is large, and the
value of the amplitude of coherent states are proportional to
$\beta^2\omega t = \beta^2 E_{ch}t/\hbar$, ($\hbar\omega = E_{ch}
\gg E_{J}$), the time for preparing an observable {\bf SS} state is
much shorter than that in other schemes.
\section{Conclusion}
In conclusion, we have presented an approach for preparing {\bf SS}
states of the mode of cavity field interacting with a
superconducting charge qubit. In contrast to other schemes we
include nonlinear effects. In general, approximations are made
directly to the full Hamiltonian in equation (\ref{H}) neglecting
all higher orders of $\beta$. In our scheme, we first apply an
unitary transformation to the Hamiltonian (\ref{H}) and make the
relevant approximations after performing the transformation. The
result obtained holds for any value of the parameter $\beta$. In the
regime in which $\hbar\omega \beta \gg E_{J}$, which can be obtained
for$\beta \ge 0.25$, the Hamiltonian becomes a squeezed Hamiltonian.
Based on the measurement of charge states, we show that {\bf SS}
states of a single-mode cavity field can be generated. Here, as
$|\beta|$ is large, and the amplitude of coherent states are
proportional to $\beta^2 E_{ch}t/\hbar$, the time for preparing
observable {\bf SS} states is much shorter than in the linear
regimes.

\end{document}